\documentclass[aps,prl,floats,twocolumn,floatfix,amsmath,showpacs,unsortedaddress,superscriptaddress]{revtex4}

\usepackage[latin1]{inputenc}
\usepackage{graphicx}

\newcommand{\be}{\begin{equation}}
\newcommand{\ee}{\end{equation}}
\newcommand{\bea}{\begin{eqnarray}}
\newcommand{\eea}{\end{eqnarray}}
\def\bma{\begin{mathletters}}
\def\ema{\end{mathletters}}

\newcommand{\ket}[1]{ | \, #1  \rangle}
\newcommand{\bra}[1]{ \langle #1 \,  |}

\newcommand{\QAv}[1]{\langle \, #1 \, \rangle}

\begin{document}

\title{Coherence Properties of Guided-Atom Interferometers}
\author{H. Kreutzmann}
\email{kreutzm@itp.uni-hannover.de}
\author{U. V. Poulsen}
\author{M. Lewenstein}
\affiliation{Institut f\"ur Theoretische Physik,
	    Universit\"at Hannover, 30167 Hannover, Germany}
\author{R. Dumke}
\author{W. Ertmer}
\author{G. Birkl}
\affiliation{Institut f\"ur Quantenoptik,
	    Universit\"at Hannover, 30167 Hannover, Germany}
\author{A. Sanpera}
\affiliation{Institut f\"ur Theoretische Physik,
	    Universit\"at Hannover, 30167 Hannover, Germany}

\date{\today}


\begin{abstract}
We present a detailed investigation of the coherence properties of
beam splitters and Mach-Zehnder interferometers for guided atoms.  It
is demonstrated that such a setup permits coherent wave packet
splitting and leads to the appearance of interference fringes.  We
study single-mode and thermal input states and show that even for
thermal input states interference fringes can be clearly observed,
thus demonstrating the multimode operation and the robustness of the
interferometer.
\end{abstract}

\maketitle

The investigation and exploitation of the wave properties of atomic
matter is of great interest for fundamental as well as applied
research and constitutes, therefore, one of the most active research 
areas in atomic physics and quantum optics.  Of special interest
is the field of atom interferometry~\cite{atominter}.  For an
interferometer, it is crucial that the beam splitters and mirrors are
coherent, i.e., they must not disturb the phase of the matter wave in
an uncontrollable fashion.  Then a phase shift in one of the paths
results in a change of the output signal, and any external influence
inducing a phase shift is, in principle, accessible to measurement. 
Compared to light interferometry, matter wave interferometry with
cold atoms offers a much higher sensitivity for certain
applications~\cite{atominter}.  Furthermore, atoms couple efficiently
to a wider variety of external interactions, thus extending the
applicability of interferometric measurements~\cite{atominter}.

A novel approach arises from the use of guided
atoms~\cite{guided,houde00,dumke02}.  Miniaturized setups for matter
wave interferometry with increased stability, large beam separation,
and large enclosed areas become
possible~\cite{hinds01,haensel01,anders02,dumke02}.  These features are
specifically appealing to atom-interferometrical measurement of
inertial forces~\cite{gyroscop} and to the investigation of
Bose-Einstein condensates in microstructures~\cite{BEC}.  Due to the
physics involved in the guiding and beam splitting processes, the
construction and analysis of guided-atom interferometers become
challenging tasks.  In order to assess the level of performance that
can be reached with realistic setups, coherence and interference,
also for mixed input states as well as non-perfect beam splitters,
have to be investigated in detail.

In this Letter we study guided-atom interferometers for neutral atoms
by solving numerically the time dependent Schrödinger equation for
realistic, experimentally accessible configurations.  Our study
addresses the main issues of atom waveguide propagation, as well as
coherence and interference using X-shaped guided-atom beam
splitters~\cite{guided,houde00,dumke02,newones}. In particular, our
calculations apply to the interferometer scheme originally proposed
in~\cite{birkl01} in which neutral atoms are guided in dipole
potentials created by micro-fabricated optical systems.  Many of the
experimental prerequisites for that proposal such as atom guides,
beam splitters and even geometries composing a complete Mach-Zehnder
interferometer for atoms have been realized already~\cite{dumke02}.

In the scheme considered here, the beam splitters are achieved by
crossing two optical waveguides at some angle~\cite{houde00,dumke02}.
 Such a beam splitter splits the atomic wave packet in coordinate
space without affecting the internal state.  Nevertheless, internal
state selective interferometry is also possible in this
scheme~\cite{dumke02}.  During the splitting process the system might
exhibit quantum reflections and tunneling between adjacent guides
and, therefore, the dynamics is in general complicated.  We will,
however, demonstrate that such a beam splitter is coherent even for a
thermal distribution of atoms with an average energy far exceeding
the level spacing of the transverse confinement.

To study the properties of the beam splitter and the interferometer
we use the Split-Operator method to solve the time dependent
Schrödinger equation
\be
i\hbar\frac{\partial \Psi(x,y,t)}{\partial t}
=\Big(-\frac{\hbar^2}{2m}\nabla^2+V(x,y)\Big)\Psi(x,y,t),
\ee
where the potential $V(x,y)$ includes the
waveguides potentials as well as any other relevant potential in the problem. 
For simplicity we shall assume throughout this paper that the atomic
wave packet is tightly confined in the third dimension so that the
dynamics is well described within a two-dimensional treatment.

The central element of any interferometer is the beam splitter. 
Consequently, we start our discussion with a detailed analysis of an
X-shaped guided-atom beam splitter (BS) created by crossing 
two identical waveguides L$_i$ and L$_j$ at an angle
$\gamma$:
$$
U_{\text{BS}}(x,y)=
U_i(x,y)+U_j(x\cos\gamma-y\sin\gamma,y\cos\gamma+x\sin\gamma)
\label{eq:beams}
$$
Each waveguide consists of a Gaussian potential of depth
$U_0$ and width $\sigma$; thus a waveguide L$_i$ along the
y-direction centered at $x_0$ is represented by: 
$$U_i(y,x)=-U_0 e^{-(x-x_0)^2/2\sigma^2}$$ 
Coupling from one guide to the other occurs for any angle 
$\gamma\neq 90^\circ$.  We observe, however, that the doubling of the
potential depth at the crossing of the two waveguides induces quantum
reflections and a highly non-adiabatic dynamics for typical initial
momenta of the atomic wave packet.  These effects are clearly
undesirable for interferometry.  For a micro-optical realization of
the waveguides, they can be easily avoided by reducing the light
intensity at the crossing through overlay of an absorptive mask or by
adding a compensating extra potential (e.g., a blue detuned laser
field) so that the depth of the total potential at the crossing is
equal to the one of each waveguide alone.  In our simulations we have
taken into account that this compensation might not be perfect.

\begin{figure}
\includegraphics[width=0.4\linewidth]{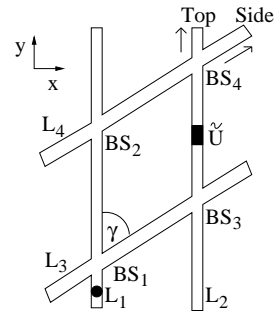}
\caption{Schematic view of a Mach-Zehnder interferometer for guided atoms
  using 4 identical waveguides crossing at an angle $\gamma$.  The
  initial atomic wave packet is represented by a dot in waveguide L$_1$
  below the first beam splitter (BS$_1$).  The location of the
  phase shift potential $\tilde U(x,y)$ is also depicted.}\label{fig:scheme}
\end{figure}

Our choice of parameters closely matches the relevant experimental
parameters of refs.~\cite{birkl01,dumke02} for $^{85}$Rb atoms guided
in dipole potentials, far detuned below the 
$5$S$_{1/2}\rightarrow 5$P$_{3/2}$ transition at 780 nm.  A typical
configuration consists of waveguides of width $\sigma = 0.54$ $\mu$m
(corresponding to a Gaussian beam with $1/e^2$ waist $w_0 = 1.1$
$\mu$m) at a laser wavelength of 830 nm and an intensity of 
$I = 1.1 \times 10^5$ W/cm$^2$ (less than 1 W of required laser
power).  The depth of potential is $U_0=75$ $\mu$K, the ground state
vibrational frequency $\omega = 160 \times 10^3$ s$^{-1}$, and the
rate of spontaneous scattering $\Gamma_{\text{S}} = 2.6$ s$^{-1}$
(thus it can be neglected in our discussion).  In our simulations, as
initial state we consider an atomic wave packet located in one of the
waveguides (L$_1$ in Fig.~\ref{fig:scheme}) at a typical distance of
2.5 $\mu$m from BS$_1$.  We perform simulations
for both single and multimode transverse occupation of the
waveguide.  In general, the initial transversal state can be
described by a thermal mixture
\begin{equation}
  \label{eq:th_distr}
  \rho=\frac{1}{Z} \sum_{n=0}^{\infty} e^{-E_n/k_{\text{B}} T} \ket{n}\bra{n},
\end{equation}
where $\ket{n}$ denotes the $n^{\text{th}}$ eigenstate of the
waveguide potential, $E_n$ denotes its energy and $Z$ ensures proper
normalization.  In the experiment~\cite{dumke02} an ultra-cold atomic
sample with temperature of $T = 20$ $\mu$K is loaded into the
waveguide from a single dipole trap.  Our set of parameters results
in a mean transverse occupation number of $\QAv{n}\simeq 16$ in the
waveguide.  For both single-mode ($\rho=\ket{n}\bra{n}$) and
multimode cases, we assume that the initial atomic wave packet has a
Gaussian profile along the longitudinal direction of the waveguide
with a mean momentum $p_y$ and a spread of $\Delta p_y$.  Values of
the mean momentum are in the range of $5-10$ recoil momenta
$p_{\text{r}}$ 
($p_{\text{r}}=\sqrt{2\hbar m\omega_{\text{r}}}$,
 $\omega_{\text{r}}= 24 \times 10^3$ s$^{-1}$),
with a momentum spread $\Delta p_y=2 p_{\text{r}}$.

\begin{figure}
\includegraphics[height=0.7\linewidth,angle=270,origin=c]{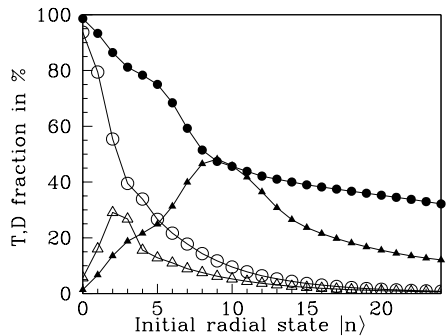}
\vspace{-1cm}
\caption{Splitting efficiency of guided-atom beam splitter: transmitted (T)
         (circles) and deflected (D) (triangles) fraction of the incoming
	 wave packet as function of the initial transverse mode of the
	 waveguide for $\gamma=45^{\circ}$.  Full symbols correspond to
	 $p_y=10p_{\text{r}}$ and open symbols to
	 $p_y=5p_{\text{r}}$.}\label{fig:norm45}
\end{figure}

The splitting of the initial wave packet depends on its initial
transversal states, its initial longitudinal momentum $p_y$, and on
the angle $\gamma$ between the guides.  In order to achieve efficient
deflection in the beam splitter, the atomic wave packet has to spend
enough time at the intersection.  Defining the crossing time as
$t_{\text{c}}=\sigma/v_y$ with $v_y=p_y/m$, efficient splitting
requires $t_{\text{c}}\gtrsim \hbar/E_n$.  For a fixed angle $\gamma$ and
a fixed initial momentum $p_y$, we observe that the dependence of the
splitting ratio on the initial transverse state $\ket{n}$ is very strong.  
This is displayed in Fig.~\ref{fig:norm45}, where we plot the fraction
of atoms transmitted towards BS$_2$ (T) and the fraction deflected
towards BS$_3$ (D) as a function of the initial transverse state for
$\gamma=45^{\circ}$ and two different initial longitudinal momenta,
$p_y=10 p_{\text{r}}$ and $p_y=5p_{\text{r}}$.  To ``count'' the
number of deflected (transmitted) atoms we use an absorbing box at
L$_3$ (L$_1$) after the beam splitter and integrate the loss of norm
in each box with time.  The transmitted and deflected fraction do not
always add up to unity.  The missing fraction consists of atoms
backscattered into the downward sections of waveguides L$_1$ and L$_3$. 
For $p_y=10p_{\text{r}}$, an approximately 50/50 splitting ratio
occurs for transverse initial states with quantum numbers 
$n\simeq 8-11$.  Losses due to backscattering are very small in this
case. For $p_y=5 p_{\text{r}}$ the optimal splitting ratio occurs for
$n \simeq 2-3$ although losses are now significantly higher.  In both
cases, the deflected fraction is narrowly peaked around its maximum
evidencing that the beam splitter acts as filter for transverse states.
This implies that even for thermal input states, efficient
splitting occurs only for a narrow group of states around
the optimal one. In both cases the fraction of deflected atoms is 
very small for the ground state ($\rho=\ket{0}\bra{0}$) of the potential.

The following simple picture helps to understand the selection of the
optimal state $\ket{n}$ for a given initial momentum $p_y$ and angle
$\gamma$: At the intersection, the wave packet is no longer confined
transversally and, therefore, expands according to its initial
transverse momentum distribution.  Optimal splitting occurs for
typical transverse momenta $p_x=\sqrt{2mE_n}$ fulfilling 
$p_x \simeq p_y \tan\gamma$.  Approximating the center of the
waveguide by a harmonic potential of frequency $\omega$ one obtains
the estimate $n_{\text{opt}}\simeq p_y^2\tan^2(\gamma)/(2m\hbar\omega)$.
Lower transverse states can be made optimal by lowering $p_y$ and/or
lowering $\gamma$.  A lower $n_{\text{opt}}$ also implies fewer states
within the peak of non-negligible splitting ratios thus enhancing the
filtering effect of the beam splitter.  Despite the simplicity of the
argument, it agrees well with our results depicted in
Fig.~\ref{fig:norm45} and with our simulations for different initial
longitudinal momenta $p_y$ and angles $\gamma$.  Only for angles
$\gamma\leq 25^{\circ}$ we find strong deviations from our estimate
since there backscattering and tunneling start to play an important
role in the splitting dynamics and the simplified argument fails to
reproduce the numerical results.

The coupling between transverse and longitudinal degrees of freedom
in the splitting process results in a complex dynamics after the beam
splitter.  We calculate the longitudinal phase profile of each wave
packet before it reaches the next beam splitter.  The transmitted
wave packet propagates along the waveguide L$_1$ almost without
distortion and its phase profile agrees well with the phase profile
of a wave packet propagating freely in a transverse harmonic
potential~\cite{borde01}.  The motion along L$_3$ is more involved. 
Although the overall behavior of the phase resembles the one of the
wave packet along L$_1$, even for a single mode initial state, the
wave packet exhibits transverse oscillatory motion of frequency
$\omega$.  This is caused by the excitation of a coherent
superposition of transverse modes and clearly
demonstrates that the splitting process is non-adiabatic.  This
oscillatory motion has to be taken into account for optimizing the
efficiency of the full Mach-Zehnder interferometer.

The full interferometer is realized by appropriately crossing four
identical waveguides.  Due to numerical limitations, we constrain 
the distance between L$_1$ and L$_2$ to about 7 $\mu$m.
~\footnote{Convergence and wave packet spreading effects were studied by
performing some runs with a waveguide separation of 21 $\mu$m.  
The results in both cases agree very well 
showing that spreading does not affect the visibility 
of the interference fringes.}
Atoms loaded in the
waveguide L$_1$ with initial momentum $p_y$ propagate to the first
beam splitter BS$_1$ and are then guided through L$_1$ and L$_3$
towards BS$_2$ and BS$_3$ respectively.  Here BS$_2$ and BS$_3$ act as
mirrors, but due to the additional output ports also establish new
loss channels.  Note that this Mach-Zehnder configuration is not
symmetric even for 50/50 beam splitters, since, as we have already
discussed, the dynamics in each arm of the interferometer becomes
quite distinct after the first beam splitter. Following 
splitting processes in the two arms are no longer equivalent.  This
lack of symmetry excludes the possibility of achieving 100\%
visibility (defined as the amplitude [i.e., $1/2$ (max-min)] of the
modulation in the top output divided by the average in the top
output) and demands optimization strategies to improve the
visibility.

\begin{figure}
\includegraphics[width=0.9\linewidth,bb=102 262 521 470]{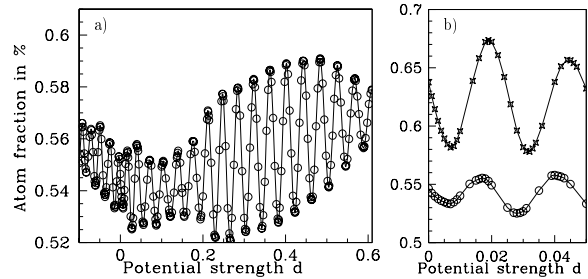}
\caption{Interference fringes for a single-mode ($\rho=\ket{0}\bra{0}$) 
  initial state versus the strength $d$ of $\tilde U(x,y)$.
  (a) Fraction of atoms (in \% of the input atom number) in the top output.
  (b) Fraction of atoms (in \%) in the top output for the initial 
interferometer size (circles) and the optimized size (L$_2$ displaced in
x-direction) (stars).}
\label{fig:ground}
\end{figure}

\begin{figure}
\includegraphics[height=0.8\linewidth,angle=270,origin=c]{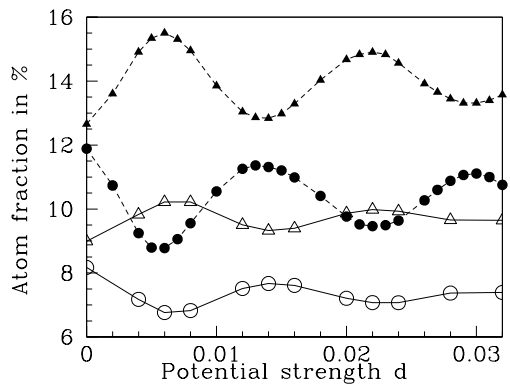}
\vspace{-1cm}
\caption{Interference for single-mode and thermal input states:
   fraction of atoms (in \%) in the side (triangles) 
   and top output (circles) versus the strength $d$ of $\tilde U(x,y)$.
   Filled symbols correspond 
  to the transverse state with optimal splitting $n=2$, open symbols
  to an initial thermal state at T=20 $\mu$K. }\label{fig:thermal} 
\end{figure}

To demonstrate that coherence is preserved during the propagation
through the full interferometer, we calculate the output signals
after the final beam splitter BS$_4$ as a function of the depth $d$
of an additional potential
$\tilde U(x,y)= -d\,U_0e^{-(x-\tilde x_0)^2/2\sigma^2} e^{-(y-\tilde y_0)^2/2\sigma^2}$ 
which is inserted in one of the arms of the interferometer (between
BS$_3$ and BS$_4$) to induce a phase shift between both arms.  This
extra potential, which can be either attractive or repulsive
depending on the sign of $d$, has a Gaussian profile along the
waveguide and smoothly lowers ($d>0$) or increases ($d<0$) the
potential depth of that part of the waveguide.  Our results are
summarized in Fig.~\ref{fig:ground} and Fig.~\ref{fig:thermal}. 
Clear periodic modulations in the number of atoms exiting each output
port (labeled top and side outputs for clarity) appear as a function
of $d$ for both single-mode (Fig.~\ref{fig:ground}) and thermal initial
state (Fig.~\ref{fig:thermal}).

For the single mode case, we choose, for simplicity, the ground state
with longitudinal momentum $p_y=10p_{\text{r}}$ as initial state.  In
Fig.~\ref{fig:ground}(a), we display the fraction of atoms at the top
output versus $d$.  The expected modulation of the atom number as a
function of $d$ is clearly visible.  We compare the period of the
oscillation with the one obtained using a simplified model of the
phase shift $\phi$ introduced by the additional potential $\tilde U$
using the classical action $S =\int \text{d}t\, \mathcal{L}$, where
$\mathcal{L}$ is the Lagrangian.  Despite the oversimplification of
this model, the calculated period is only a few \% too small for low
values of $d$.  For larger $d$ the increase in the oscillation period
is more accurately described using a WKB approximation.

The combined atom number of both outputs of the interferometer is,
for this case, small ($\approx$ 1\% of the initial atom number)
since, as shown in Fig.~\ref{fig:norm45}, the ground state splits
very inefficiently and most of the atoms leave the
interferometer at BS$_2$ and BS$_3$.  The visibility of the fringes
is also low ($\approx$ 2\%).  Better visibility can be achieved by
choosing as the initial state the one with the optimal splitting
ratio (c.f. Fig.~\ref{fig:thermal}) and using an optimized geometry. 
An optimized geometry aims at equalizing as much as possible the
contributions from both arms of the interferometer to the final
signal.  This can be accomplished by choosing the distance between
the waveguides L$_1$ and L$_2$ such that the wave packet that
propagates in L$_3$ with transverse oscillations enters the beam
splitter BS$_3$ with a mean transverse momentum appropriate to
maximize the fraction of atoms deflected towards BS$_4$.  This
optimization immediately leads to a visibility of $\approx$
7\% as shown in Fig~\ref{fig:ground}(b).

Experimentally, a single transverse mode wave packet is difficult 
to achieve.  A more realistic scenario corresponds 
to an initial thermal occupation of the transverse modes 
(Eq.~\ref{eq:th_distr}).  For this case, we
calculate the final output signal as a classical (Boltzmann) weighted
average of the signal obtained for each transverse mode $\ket{n}$
separately.  In our calculations, we now use an optimized geometry
but we consider, nevertheless, that the compensation of the double
potential in the beam splitters is not perfect allowing for a 5\%
mismatch in potential depth at the beam splitter.  
In Fig.~\ref{fig:thermal} we display the output signal 
for an initial state corresponding to all atoms in the optimal spliting 
state $\rho=\ket{2}\bra{2}$ (c.f.  Fig.~\ref{fig:norm45}) 
and for a thermal (multimode) state corresponding to T=20 $\mu$K . In both
cases $p_y = 5 p_{\text{r}}$ and $\gamma=45^{\circ}$. 
Since the beam splitter is state selective, interference
fringes are mostly due to states with a good splitting ratio as
evidenced by the fact that maxima and minima appear approximately at
the same positions in both cases.  The combined atom number for the
optimal state is about 24\% of the initial atom number, 
with visibilities up to 15\%. (Higher visibilities of $\approx$ 23\% 
can be obtained for $\rho=\ket{1}\bra{1}$ but with a much lower
signal). For the thermal initial state the different splitting
ratios corresponding to the different transverse states lower 
the total output signal
to $\approx$ 17\% but with visibilities up to 10\%.  This
clearly demonstrates the persistence of coherence and interference
even for thermal input states and allows for the operation of the
interferometer as a multimode device.  In fact, compared to the zero
temperature case (ground state), a thermal state which inherently
contains the optimal state will dramatically improve the performance
of the interferometer for large $\gamma$.

In summary, we have shown that guided-atom configu\~rations allow for
significant extensions of matter wave interferometry.  An X-shaped
beam splitter created by crossing two identical waveguides (with the
doubling of the potential properly compensated) preserves coherence
and acts analogously to a ``color filter'', selecting only a few
optimal radial states which contribute to the final signal.  In this
way coherence is preserved throughout the interferometer.  As a
consequence, temperature is not necessarily a limitation for the
observation of interference fringes but can rather be a requisite to
ensure a good performance.  Our simulations show that an additional
potential of variable strength in one of the arms of the
interferometer can produce a straightforward proof for coherence and
interference.  Finally, we have also demonstrated improvements by
applying optimization strategies.

This work is supported by the \textit{Deutsche
Forschungsgemeinschaft} (SFB\,407), the RTN Cold Quantum Gases and
the IST-project ACQP of the \textit{European Commission}.

\end{document}